\documentstyle[aps,prl,multicol,epsfig,amssymb]{revtex}

\begin{document}

\title{Two-stage Kondo effect in a quantum dot at high magnetic field}

\author{W.G. van der Wiel$^1$, S. De Franceschi$^1$, J.M. Elzerman$^1$, S. Tarucha$^2$ and L.P. Kouwenhoven$^1$}
\address{$^1$Department of Applied Physics, DIMES, and ERATO Mesoscopic
Correlation Project,\\Delft University of Technology, PO Box 5046,
2600 GA Delft, The Netherlands}
\address{$^2$ERATO Mesoscopic Correlation Project, University of Tokyo, 7-3-1, Hongo, Bunkyo-ku, Tokyo 113-0033, Japan}

\author{J. Motohisa, F. Nakajima and T. Fukui}
\address{Research Center for Integrated Quantum Electronics, Hokkaido
University, North 13, West 8, Sapporo 060-8628, Japan}

\maketitle

\begin{abstract}
We report a strong Kondo effect (Kondo temperature $\sim$ 4K) at
high magnetic field in a selective area growth semiconductor
quantum dot. The Kondo effect is ascribed to a singlet-triplet
transition in the ground state of the dot. At the transition, the
low-temperature conductance approaches the unitary limit. Away
from the transition, for low bias voltages and temperatures, the
conductance is sharply reduced. The observed behavior is compared
to predictions for a two-stage Kondo effect in quantum dots
coupled to single-channel leads.
\end{abstract}


\begin{multicols}{2}
%Introduction
The observation of the Kondo effect in quantum dots
\cite{GG98,Sara98,Schmid98} has lead to an increased experimental
and theoretical interest in this many-body phenomenon. Unlike the
conventional case of bulk metals containing magnetic impurities
\cite{Kondo64}, quantum dots \cite{Kouwenhoven97} offer the
possibility to study the Kondo effect at the level of a single
artificial magnetic impurity \cite{Glazman88}, allowing to tune
different parameters. Experiments on quantum dots have also
revealed novel Kondo phenomena that have no analog in bulk-metal
systems. In particular, multi-level Kondo effects have been
studied both theoretically
\cite{Inoshita93,Pustilnik00,Eto00,PustilnikST,PustilnikPRB,PustilnikReview}
and experimentally \cite{Sasaki00,Nygard00,Schmid00} that differ
substantially from the ordinary case of a spin-1/2 Anderson
impurity.\\
\indent In this Letter, we present results on a strong Kondo
effect in a lateral quantum dot at high magnetic field. We
associate the Kondo effect with a magnetically induced crossing
between a spin-singlet and a spin-triplet ground state
\cite{Eto00,PustilnikST,PustilnikPRB,PustilnikReview,Sasaki00,Nygard00}.
In contrast to the results for a vertical semiconductor quantum
dot \cite{Sasaki00} and for a carbon nanotube dot \cite{Nygard00},
we find a sharp reduction of the conductance at low bias voltage,
$V_{SD}$, and temperature, $T$. We ascribe the different behavior
to the number of channels in the leads which couple to the states
in the dot. In lateral dots, tunnel barriers are obtained by
successively pinching off the propagating channels. Coulomb
blockade develops when the last channel is nearly pinched off.
Therefore, only one channel in each lead is coupled to the dot
\cite{Pustilnik01}. In vertical dots however, the tunnel barrier
characteristics are determined by the growth parameters, i.e. by
the thickness of the different semiconductor materials forming the
heterostructure and their relative conduction band offsets. In
this case, more than one conducting channel can effectively couple
to the dot states. The same is true for carbon nanotubes connected
to metal leads. Our results, in combination with previous findings
\cite{Sasaki00,Nygard00}, show that screening of higher spin
states ($S \geq 1$) depends strongly on the number of channels
coupled to the (artificial) magnetic impurity. Comparison is made

\begin{figure}[htbp]
  \begin{center}
  \centerline{\epsfig{file=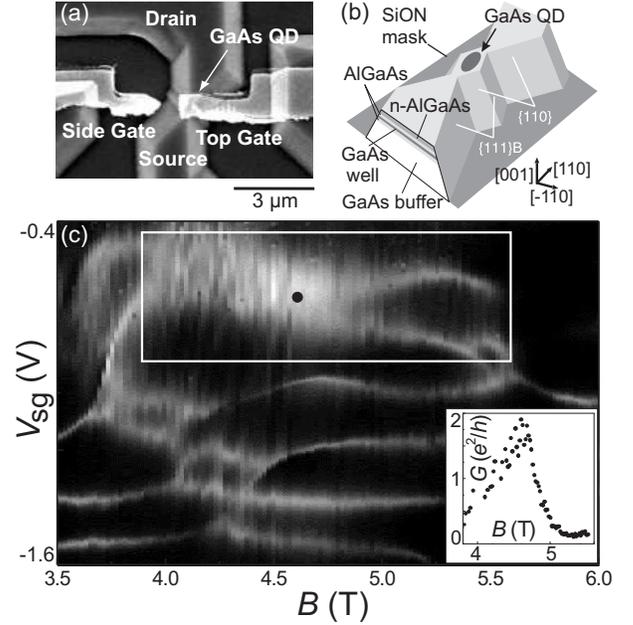, width=8cm, clip=true}}
    \caption{(a) Scanning electron micrograph. The device consists of a
    dual-gated single-electron transistor fabricated by selective area growth using metal-organic vapor
    phase epitaxy. (b) Schematic picture of the quantum dot (QD). An n-doped AlGaAs/GaAs heterostructure is selectively grown
    using a SiON mask, determining the dot shape. The crystal axes are
    indicated. The GaAs quantum well is 15 nm thick, lies 60 nm
    below the surface and has an electron density of $8.7 \times10^{15}$
    m$^{-2}$. (c) Gray-scale plot of the linear
    conductance, $G$, versus side gate voltage, $V_{sg}$, and magnetic
    field, $B$. The top gate voltage is fixed at -266 mV. Light gray lines
    indicate Coulomb peaks. The dark regions correspond to Coulomb blockade.
    The white window encloses a region of enhanced valley conductance,
    which is the focus of the present study. (inset) $G$ versus $B$ in the middle of the Coulomb valley
    within the white window.}
  \end{center}
  \label{fig1}
\end{figure}

\noindent to recent theoretical studies on quantum dots in Refs.
\cite{Pustilnik01,Hofstetter}, which are
partly inspired on the experimental work presented here.\\
%Device
\indent Our device (Fig$.$ 1a,b) consists of a lateral quantum dot
\cite{Nakajima01} with a nominal diameter $\sim$300 nm that is
further decreased by application of a negative voltage to the top
gate electrode (Fig$.$ 1a). The top gate is also used to tune the
tunnel barriers between the dot and the source and drain leads. A
side gate electrode is used to change the electrostatic potential
on the dot, although an effect on the dot shape and tunnel barrier
characteristics is unavoidable. Basic characteristics of the
device are reported in Refs. \cite{MotohisaPhysE,MotohisaAPL}. All
measurements have been performed in a dilution refrigerator with a
base temperature of 15 mK, using a standard lock-in technique with
an ac voltage between source and drain of 5 $\mu$V. \\
%Results
\indent Figure 1c shows the linear conductance, $G$, through the
dot versus side gate voltage, $V_{sg}$, and perpendicular magnetic
field, $B$. The $B$-dependence of the Coulomb peaks (light gray
lines) is rather complicated and non-monotonic. In some regions an
enhanced valley conductance is observed. We will focus on the
region within the white window. Moving from right to left in this
window, two Coulomb peaks approach each other and the valley
conductance increases. The inset to Fig$.$ 1c shows $G$ versus $B$
in the middle of the Coulomb valley. The highest conductance is
achieved around $B$ = 4.6 T and $V_{sg}$ = -625 mV (indicated by a
"{\large $\bullet$}" in Fig$.$ 1c). This local enhancement of the
valley conductance is ascribed to a singlet-triplet (S-T) Kondo
effect, as we substantiate below.

\begin{figure}[htbp]
  \begin{center}
    \centerline{\epsfig{file=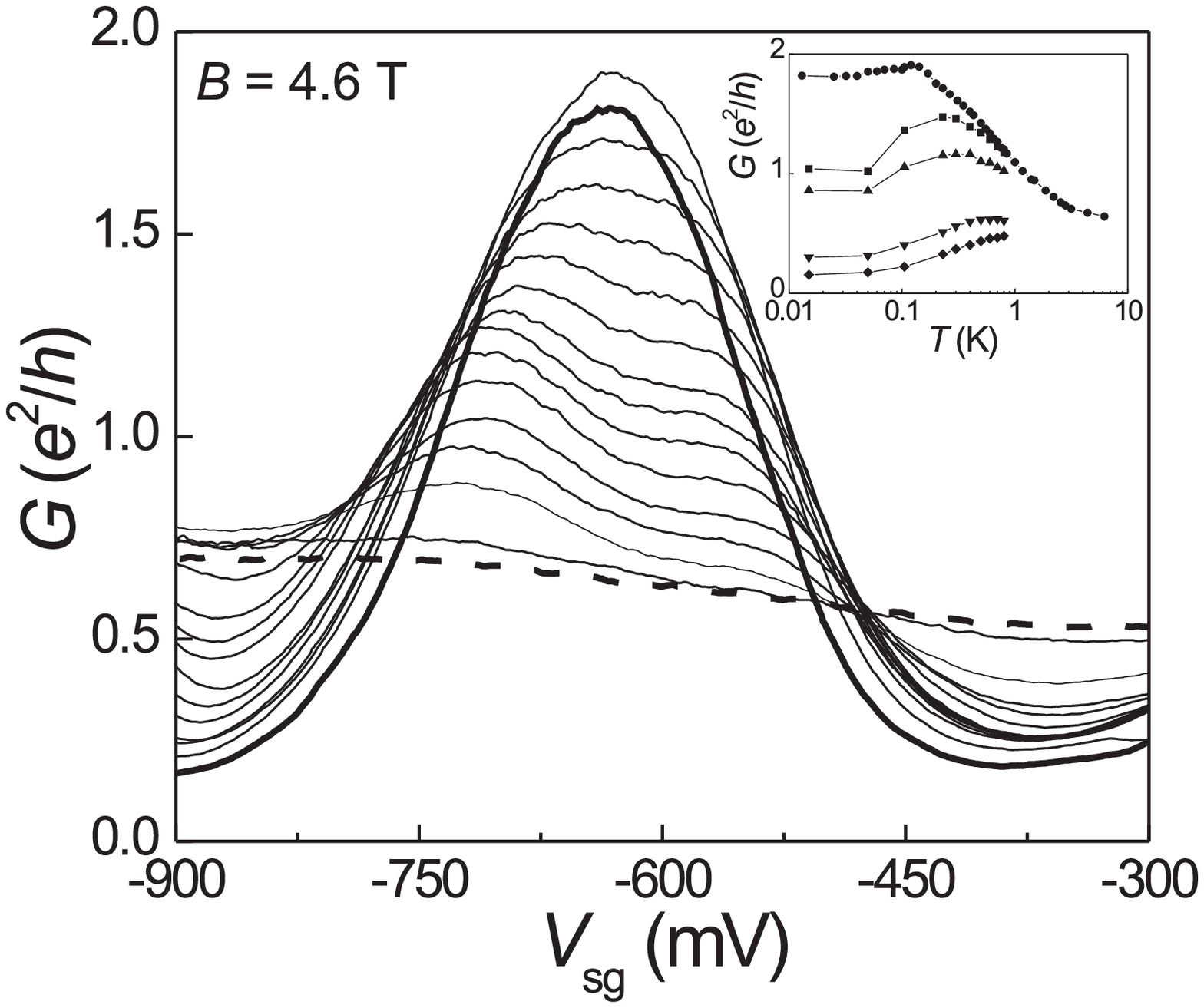, width=7.5cm, clip=true}}
    \caption{$G$-$V_{sg}$ traces taken at $B$ = 4.6 T and different
    temperatures, $T$, between 15 mK (thick solid line) and 6 K (thick dashed line).
    At the lowest $T$, $G$ approaches the unitary limit at $2e^2/h$. (inset) $G$ versus $T$ in the middle of the
    Coulomb valley ($V_{sg}$ = -625 mV) for different $B$ ($\blacksquare$ at 4.4 T, {\Large $ \bullet $} at 4.6 T,
    $\blacktriangle$ at 4.8 T, $\blacktriangledown$ at 5.0 T, $\blacklozenge$ at 5.2 T). At the degeneracy field
    $B = B_0$ = 4.6 T, $G$ decreases logarithmically between
    0.1 and 2 K. The curves for other $B$ show a clear bump that
    shifts to higher $T$ with increasing distance to $B = B_0$.}
    \label{fig2}
  \end{center}
\end{figure}

To illustrate the Kondo character of the enhanced valley
conductance, we show in Fig$.$ 2 the $T$-dependence of $G$ at $B$
= 4.6 T (which we will call $B_0$). At low $T$, $G$ at $V_{sg}$ =
-625 mV approaches the unitary limit at $2e^2/h$
\cite{Ng88,Kawabata91,Wiel00}. Around this side gate voltage, $G$
decreases with increasing $T$. The upper curve in the inset to
Fig$.$ 2 shows the $T$-dependence of $G$ in the middle of the
Coulomb valley. The observed behavior is a typical signature of
Kondo correlations. Up to our highest $T$ of 6 K, the Kondo valley
conductance keeps on decreasing, indicating an unusually strong
Kondo effect. This is quite remarkable, considering the large
magnetic field.

\begin{figure}[htbp]
  \begin{center}
    \centerline{\epsfig{file=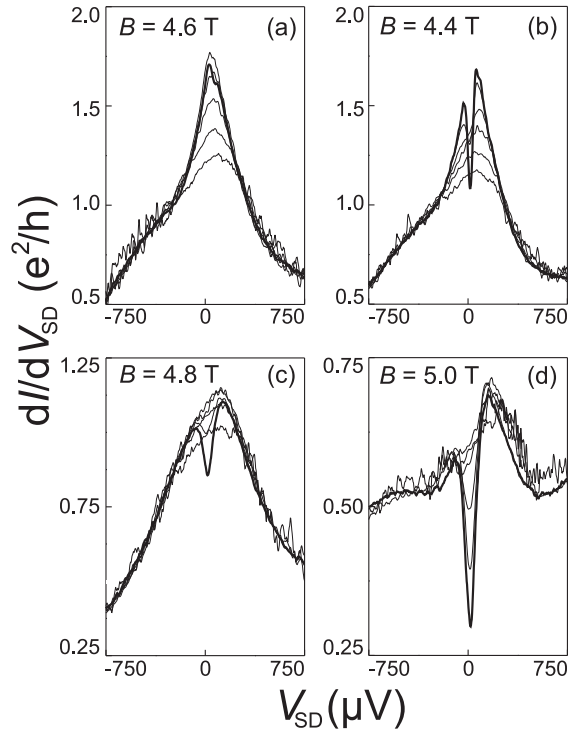, width=7.5cm, clip=true}}
    \caption{(a)-(d) $T$-dependence (between 15 mK, solid line, and
    800 mK) of the differential conductance, $dI/dV_{SD}$, versus source drain
    voltage, $V_{SD}$, in the middle of the Coulomb valley for different $B$.}
    \label{fig3}
  \end{center}
\end{figure}
\vspace{-1cm}

In Fig$.$ 3a we show the $T$-dependence of the differential
conductance, $dI/dV_{SD}$, versus source-drain voltage, $V_{SD}$
measured at $B_0$ and $V_{sg}$ = -625 mV. In the ordinary,
spin-1/2 Kondo effect, the Kondo resonance in the
$dI/dV_{SD}$-$V_{SD}$ characteristics should split by $2 |g| \mu_B
B$ \cite{splitting}, where $g$ is the Land\'{e} factor and $\mu_B$
the Bohr magneton.  Assuming $g = -0.44$, as in bulk GaAs, this
splitting is 230 $\mu$eV at $B_0$. However, this is not observed
in the $dI/dV_{SD}$-$V_{SD}$ curve at $B_0$ (thick solid line in
Fig$.$ 3a). Instead, we measure a single zero-bias resonance with
a full width at half maximum (FWHM) of about 300 $\mu$eV,
corresponding to a Kondo temperature $T_K \sim$ FWHM$/ k_B
\approx$ 4 K in the center of the Coulomb valley. This Kondo
temperature is considerably larger than the values found in
earlier experiments
\cite{GG98,Sara98,Schmid98,Sasaki00,Nygard00,Wiel00}.\\
\indent Based on the above considerations, we believe that the
explanation for the enhanced $G$ is an S-T transition around
$B_0$, assuming an even number of confined electrons. As a result
of the generalized Hund's rule \cite{Tarucha00}, the nature of the
ground state near a level crossing can change from a singlet to a
triplet, or vice versa. An S-T degeneracy leads to an enhanced
low-temperature conductance as shown experimentally in Ref.
\cite{Sasaki00,Nygard00} and theoretically in Refs.
\cite{Eto00,PustilnikST}. The local conductance enhancement at
$B_0$ in the present study suggests a similar S-T Kondo effect. In
the vertical quantum dot studied in Ref. \cite{Sasaki00}, it is
possible to determine the number of electrons in the dot and to
identify the singlet and triplet states. The present data do not
allow us to unambiguously identify the spin character of the
states on both sides of $B_0$.

\begin{figure}[htbp]
  \begin{center}
    \centerline{\epsfig{file=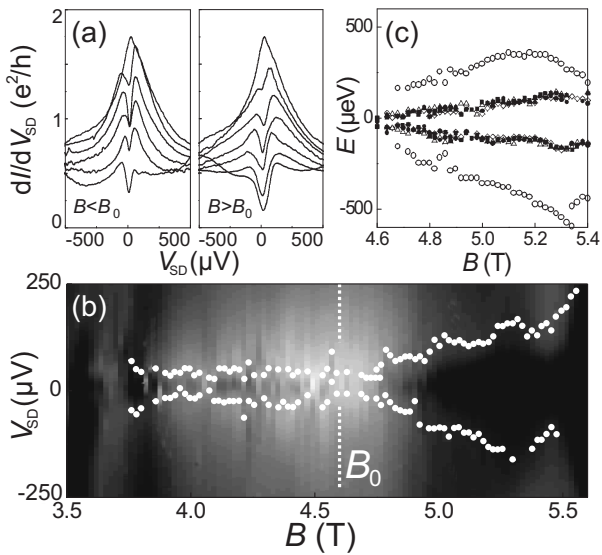, width=8cm, clip=true}}
    \caption{(a) $dI/dV_{SD}$ versus $V_{SD}$ between 3.8 and 4.6 T
    (bottom to top, left), and between 4.6 and 5.2 T (top to
    bottom, right). (b) Gray-scale plot of $dI/dV_{SD}$ versus $B$ and
    $V_{SD}$. The edges of the zero bias dip are indicated by white dots.
    (c) Comparison between the $B$-evolution
    of the dip edges (different symbols correspond
    to different $V_{sg}$-values within the Coulomb valley) and that of the Coulomb peaks ({\Large
    $\circ$}). Note that the distance between the Coulomb peaks
    includes a (constant) charging energy.}
    \label{fig4}
  \end{center}
\end{figure}
\vspace{-0.75cm}

An interesting and unexpected feature is observed when moving away
from $B_0$. A sharp dip develops around $V_{SD}$ = 0 within a
broader Kondo resonance. The dip vanishes upon increasing $T$,
restoring the broad Kondo resonance. This is shown in Figs$.$ 3b-d
for the representative fields $B$ = 4.4 T, 4.8 T and 5.0 T. The
broad Kondo resonance disappears at a much slower rate with $T$
than the sharp dip, leading to the non-monotonic behavior in the
corresponding $G$-$T$ curves in the inset to Fig$.$ 2. Moving away
from $B_0$, the position of the maximum in $G$ shifts to
higher $T$ and the low-$T$ conductance decreases.\\
\indent Increasing $B$ from $B_0$, we find that the dip width
grows up to $\sim$250 $\mu$V (see Figs$.$ 4a,b). In Fig$.$ 4c we
compare the position of the Coulomb peaks (converted from $V_{sg}$
to energy \cite{alpha}) to the positions of the dip edges. We
expect the on-site Coulomb interaction energy to be essentially
constant within the window in Fig$.$ 1c. Hence, we believe that
the $B$-evolution of the Coulomb peak spacing reflects the
$B$-dependence of the energy difference between two consecutive
levels of the dot, i.e. the S-T energy spacing. The dip width
shows qualitatively the same $B$-dependence as the spacing between
the Coulomb peaks, and thus the same as that of the S-T energy
spacing. The dip width at fixed $B$ is independent of $V_{sg}$
within the Coulomb valley where the $dI/dV_{SD}$-$V_{SD}$
characteristics are measured (see Fig$.$ 4c). This observation is
in line with a relation between the dip width and the level
spacing, irrespective of the absolute level energies.\\
\indent For $3.8 < B < B_0$ we find a dip whose width does not
vary significantly with $B$ (see Figs$.$ 4a,b). From Fig$.$ 1c we
note that, in the same $B$-range, the Coulomb peaks are closely
and constantly spaced, denoting a virtually constant $\Delta$.
This may explain why there is no clear $B$-dependence of the dip
width for $B < B_0$.\\
\indent In summary, on both sides of $B_0$ we observe a
non-monotonic $T$-dependence of $G$ and a narrow anti-resonance
superimposed on a broader Kondo resonance. The widths of these
resonances correspond to two clearly distinct energy scales. Our
observations are in qualitative agreement with recent theoretical
predictions for quantum dots coupled to single-channel leads
\cite{Pustilnik01,Hofstetter}. The single-channel character of the
leads results in both studies, but under different conditions, to
a {\it two-stage} Kondo effect. Two-stage Kondo phenomena have
been theoretically studied also in other systems, such as coupled
magnetic impurities in metals \cite{2stageinmetals}. However, no
experimental observation has been reported so far.

%explanation
For a two-stage Kondo effect to occur in quantum dots, it is
necessary that the dot has a spin state $S \geq 1$ coupled to
single-channel leads, as illustrated in Fig$.$ 5a for $S = 1$.
Instead of one low-energy scale, $T_K$, a two-stage Kondo effect
is characterized by two separate energy scales $T_{K1}$ and
$T_{K2}$ ($T_{K2} < T_{K1}$). The first-stage screening process
with characteristic energy scale $T_{K1}$ is an {\it
underscreened} Kondo effect, reducing the net spin from $S = 1$ to
$S = 1/2$. The second stage of the Kondo effect, with a smaller
energy scale $T_{K2}$, reduces the spin to $S = 0$, forming a spin
singlet. For $T_{K2} < (T,V_{SD}) < T_{K1}$, the first-stage Kondo
screening overcomes the Coulomb blockade and $G$ is expected to
reach the unitary limit. For $(T,V_{SD}) < T_{K2}$, the second
stage of the Kondo effect quenches the first one, resulting in a
suppression of $G$, ideally to zero. A finite electron temperature
or Zeeman energy may reduce the effect of the second screening
process. The signatures of a two-stage Kondo effect are a
non-monotonic $T$-dependence of $G$ and a sharp dip superimposed
on the usual zero-bias resonance in the $dI/dV_{SD}$,

\begin{figure}[htbp]
  \begin{center}
    \centerline{\epsfig{file=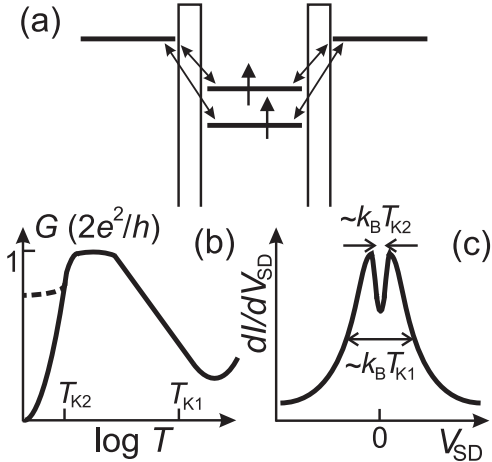, width=6.5cm, clip=true}}
    \caption{Schematics of a two-stage Kondo effect. (a) Two single-particle states
    with total spin $S = 1$ coupled to single channel leads. (b) $T$-dependence of $G$
    showing the different regimes characterized by the energy scales $T_{K1}$ and $T_{K2}$.
    The dashed line corresponds to the case of a finite electron temperature and/or Zeeman
    energy. (c) $dI/dV_{SD}$-$V_{SD}$ showing a dip with a width $\sim$$k_B T_{K2}$ within a
    Kondo resonance with width $\sim$$k_B T_{K1}$.}
    \label{fig5}
  \end{center}
\end{figure}
\vspace{-0.8cm}

\noindent as schematically depicted in Figs. 5b and 5c, respectively.\\
\indent Our results are very similar to the schematic graphs in
Fig$.$ 5b,c, supporting an interpretation in terms of a two-stage
Kondo effect. As the character of the ground state on both sides
of $B_0$ may differ, the mechanism underlying the two-stage Kondo
effect need not be the same though. Hofstetter and Schoeller find
a two-stage Kondo effect for a singlet ground state with a nearby
triplet excited state \cite{Hofstetter}. Pustilnik and Glazman
predict a two-stage Kondo effect for an $S \geq 1$ ground state
when there is significant asymmetry in the dot-lead coupling
\cite{Pustilnik01}. The latter model, however, does not include
dot excitations, which
can be important in proximity of an S-T transition.\\
\indent For $B > B_0$, we can tune the S-T spacing, $\Delta$,
between $\sim$0 and a few hundred $\mu$eV (see open white circles
in Fig$.$ 4c). Due to the high Kondo temperature ($k_B T_{K1} \sim
300 \mu$eV), we cover the range from $\Delta \ll k_B T_{K1}$ to
$\Delta \approx k_B T_{K1}$. Over a $B$-range of about 0.5 T we
observe that the dip width scales linearly with $\Delta$. From the
observed dip width (Fig$.$ 4c) we can estimate $T_{K2} \lesssim$ 1
K. The observed behavior for $B > B_0$ is in line with the results
of Ref. \cite{Hofstetter} for the case of a singlet ground state
with a nearby triplet. There the energy scale $T_{K2}$ increases
linearly with $\Delta$ for $\Delta \sim k_B T_{K1}$. We also note
that the absence of a triplet Kondo resonance for $B
> B_0$ favors the interpretation of a singlet ground state here.\\
\indent For $B < B_0$ there is no clear $B$-dependence of the
conductance dip. Possibly, a singlet ground state is close to a
triplet excited state over a relatively wide $B$-range. In this
case an explanation as for $B > B_0$ applies. This interpretation
requires a transition from a singlet ground state ($B < B_0$) to a
region of S-T degeneracy ($B \sim B_0$) to again a singlet ground
state ($B > B_0$). Alternatively, the ground state for $B < B_0$
could also be a triplet, in which case the mechanism proposed in
Ref. \cite{Pustilnik01} may be responsible for the dip.\\
\indent We conclude that, irrespective of the precise model, our
results can be interpreted in terms of a two-stage Kondo effect,
characterized by two well-separated energy scales.\\
%acknowledgements
\indent We thank W. Hofstetter, H. Schoeller, M. Pustilnik, L.
Glazman, T. Costi, W. Izumida, M. Eto and R. Schouten for their
help. We acknowledge financial support from the Specially Promoted
Research Grant-in-Aid for Scientific Research; Ministry of
Education, Culture, Sports, Science and Technology in Japan; the
Dutch Organization for Fundamental Research on Matter; the Core
Research for Evolutional Science and Technology (CREST-JST); and
the European Union through a Training and Mobility of Researchers
Program network. 

\end{multicols}

\begin{references}
\bibitem{alpha} We use $\varepsilon = \alpha V_{sg}$ + constant,
with $\alpha = C_{sg}/C_{\Sigma}$ = 3.4 $\mu$eV/mV. Here,
$\varepsilon$ is the level energy, $C_{sg}$ the capacitance of the
side gate and $C_{\Sigma}$ the total capacitance of the dot.
\end{references}
\end{document}